\documentclass[proceedings]{JHEP} 

\newcommand {\cI}{{\cal I}}
\newcommand {\cJ}{{\cal J}}

\newcommand {\cL}{{\cal L}}

\newcommand {\cN}{{\cal N}}
\newcommand {\cO}{{\cal O}}

\newcommand{\bP}{{\bf P}}

\newcommand{\bR}{{\bf R}}

\newcommand{\bX}{{\bf X}}

\newcommand{\bZ}{{\bf Z}}
\def\a{\alpha}

\def\b{\beta}

\def\d{\delta}

\def\g{\gamma}

\def\j{\psi}

\def\o{\omega}
\def\p{\pi}
\def\q{\theta}

\def\s{\sigma}

\def\x{\xi}

\def\D{\Delta}

\def\L{\Lambda}

\def\Q{\Theta}

\newcommand{\ad}{{\dot{\alpha}}}                           
\newcommand{\bd}{{\dot{\beta}}}                            
\newcommand{\ve}{\varepsilon}                            

\newcommand{\pa}{\partial}                           
\newcommand{\hf}{\frac12}
\newcommand{\abar}{\bar{a}}
\newcommand{\bbar}{\bar{b}}
\newcommand{\cbar}{\bar{c}}

\newcommand{\be}{\begin{equation}}
\newcommand{\ee}{\end{equation}}
\newcommand{\bea}{\begin{eqnarray}}
\newcommand{\eea}{\end{eqnarray}}
\newcommand{\non}{\nonumber}
 
\conference{Quantum aspects of gauge theories, supersymmetry and
unification }

\title{Implications of \mbox{$\cN$} = 2 
Superconformal Symmetry}

\author{Sergei Kuzenko, Stefan Theisen
\\
Sektion Physik der
Ludwig-Maximilians-Universit\"at \\
Theresienstra{\ss}e 37, D-80333 M\"unchen, Germany
\\
        E-mail: \email{sergei, 
theisen@theorie.physik.uni-muenchen.de }}

\abstract{We review recent results on
the general structure of two- and three- 
point functions of the supercurrent
and the flavor current of $\cN=2$ 
superconformal field theories.}

\begin{document}

  \section{Introduction}
This note is a brief review of the results obtained
in our recent paper \cite{KT} where we analysed 
the general structure of two- and three- 
point functions of the supercurrent
and the flavor current of $\cN=2$ 
superconformal field theories. 
Our research was inspired by (i)
similar results obtained by Osborn 
\cite{osborn}
for $\cN=1$ superconformal field theories;
(ii) $\cN$--extended superconformal kinematics
due to Park \cite{park2}, 
in particular the existence of nilpotent superconformal
invariants of three points; (iii)
the conjecture of Maldacena \cite{maldacena}
(see \cite{aharonyetal} for a review), 
which relates superconformal gauge theories in four dimensional 
Minkowski space to extended gauge supergravities 
in five dimensional anti-de-Sitter space. 

In $\cN=1$ superconformal field theory, 
the conserved currents are contained in two 
different supermultiplets: (i) the supercurrent
$J_{\a \ad}$ \cite{fz} containing 
the energy-momentum tensor $\Theta_{mn}$,
the supersymmetry currents
$j_{m \hat{\a}}$ ($\hat{\a} = \a , \ad $) 
and the axial current $j^{(R)}_m$;
(ii) the flavor current multiplet $L^{\bar{a}}$ 
\cite{WZ} containing
the conserved flavor current $v^{\bar{a}}_m$ among 
its components. Both $J_{\a \ad}$ and $L^{\bar{a}}$
are real $\cN=1$ superfields, and they satisfy the 
conservation equations
\bea
{\bar D}^\ad J_{\a \ad} &=& 
D^\a J_{\a \ad} 
= 0~, \\
{\bar D}^2 L^{\bar{a}} &=&
D^2 L^{\bar{a}}
= 0~.
\eea

In $\cN=2$ superconformal field theory, 
the conserved currents are contained in two 
different supermultiplets: (i) the supercurrent
$\cJ$ \cite{sohnius,hst} whose components include
the energy-momentum tensor $\Theta_{mn}$,
the SU(2) $R$-current $j^{(ij)}_m$ ($i,j= 
\underline{1},\underline{2}$), the axial current
$j^{(R)}_m$ and the $\cN=2$ supersymmetry currents
$j^i_{m \hat{\a}}$;  (ii) the flavor current 
multiplet $\cL^{\bar{a}}_{ij}$ \cite{Mez,hst} containing
the conserved flavor current $v^{\bar{a}}_m$ among 
its components. Both $\cJ$ and $\cL^{\bar{a}}_{ij}$
are real $\cN=2$ superfields 
($\overline{ \cL_{ij} }= \cL^{ij}$), and they satisfy the 
conservation equations
\bea
D^{ij} \cJ &=& {\bar D}^{ij} \cJ 
= 0~, \label{scc2}  \\
D^{(i}_\a \,\cL^{jk)} 
&=& {\bar D}^{(i}_\ad \,\cL^{jk)} 
= 0~,
\label{scc2f}
\eea 
where $D^{ij} = D^{\a ( i} D^{j)}_\a$, ${\bar D}^{ij}
= {\bar D}^{(i}_\ad {\bar D}^{j) \ad }$. 

Any $\cN=2$ superconformal field theory
is a special $\cN=1$ superconformal model.
Therefore, it is useful to know
the decomposition of $\cJ$ and $\cL_{ij}$
into $\cN=1$ multiplets. For that purpose
we introduce the $\cN=1$ spinor covariant derivatives 
$D_\a \equiv D^{\underline{1}}_\a$, ${\bar D}^\ad \equiv
{\bar D}^\ad_{\underline{1}}$ and define the $\cN=1$ projection 
$U| \equiv U(x,\q^\a_i, {\bar \q}^j_\ad ) 
|_{ \q_{\underline{2}} = 
{\bar \q}^{\underline{2}} = 0}$ of an arbitrary $\cN=2$ superfield
$U$. It follows from (\ref{scc2}) that 
$\cJ$ is composed of three independent $\cN=1$ 
multiplets
\bea
J  & \equiv & \cJ|~, \qquad  J_\a  \equiv 
D^{ \underline{2} }_\a \cJ |~, 
\label{n=1s-ccomponents} \\
J_{\a \ad} & \equiv & \hf \,
[D^{ \underline{2} }_\a  ,  
{\bar D}_{\ad \underline{2} }] \cJ| 
-\frac{1}{6} \, [D^{ \underline{1} }_\a  , 
{\bar D}_{\ad \underline{1} }] \cJ| ~,
\non
\eea
while the $\cN=1$ flavor current multiplet 
is identified as follows
\be
L \equiv {\rm i} \, \cL^{ \underline{12}}|~. 
\ee
Here $J$ and $J_\a$ satisfy the conservation equations
\bea
{\bar D}^2 J &=&
D^2 J= 0~, \\
D^\a J_\a &=& {\bar D}^2 J_\a = 0 ~.
\eea
The spinor object $J_\a$ contains the second supersymmetry 
current and two of the three 
SU(2) currents, namely those
which correspond to the symmetries belonging to SU(2)$/$U(1).
Finally, the scalar $J$ contains the current corresponding 
to the special combination of the $\cN=2$ U(1) $R$-transformation 
and SU(2) $\s_3$-rotation which leaves $\q_{\underline{1}}$ 
and ${\bar \q}^{\underline{1}}$ invariant.

\section{Superconformal building blocks} 
 
In $\cN$--extended global superspace 
$\bR^{4|4\cN}$ parametrised by 
$z^A = (x^a, \q^\a_i, {\bar \q}^i_\ad) $,
an infinitesimal superconformal transformation
\bea
& z^A ~ \longrightarrow ~ z^A ~+~ \x 
\cdot z^A~, & \\
& \x = {\overline \x} = \x^a (z) \pa_a + \x^\a_i (z)D^i_\a
+ {\bar \x}_\ad^i (z) {\bar D}^\ad_i & \non 
\eea
is generated by a superconformal Killing vector $\x$ 
defined to satisfy 
\be
[\x \;,\; D^i_\a ] \; \propto \; D^j_\b ~.
\ee   
{}From here it follows
\be
\x^\a_i = -\frac{\rm i}{8} {\bar D}_{\bd i} \x^{\bd \a}\;, \qquad
{\bar D}_{\bd j} \x^\a_i = 0
\label{spinsc}
\ee
while the vector component of $\x$ is constrained by
\bea
& D^i_{(\a} \x_{\b )\bd} = {\bar D}_i^{(\ad} \x^{\bd ) \b }=0~,& \\
& \Longrightarrow   \quad
\pa_a \x_b + \pa_b \x_a = \hf\, \eta_{ab}\, \pa_c \x^c\;. & \non 
\eea 
{}For $\cN < 4$, the algebra of superconformal Killing vectors
is isomorphic to the $\cN$--extended superconformal algebra,
su$(2,2|\cN)$.

Let us introduce the parameters of generalized 
Lorentz $\o_{(\a \b )}$,   
scale--chiral $\s$ 
and SU$(\cN)$ transformations $\L_i{}^j$ 
($\L^\dag -  \L = {\rm tr}\; \L = 0$) generated by $\x$
\bea
[\x \;,\; D^i_\a ] &=& - (D^i_\a \x^\b_j) D^j_\b \non \\
&=& \o_\a{}^\b  D^i_\b   -{\rm i} \L_j{}^i \; D^j_\a  \non \\
&- & \frac{1}{\cN}
\Big( (\cN-2) \s + 2 {\bar \s}  \Big) D^i_\a~.
\eea
A {\it primary} superfield $\cO (z)$,
carrying some number of undotted and dotted
spinor indices
and transforming in some
representation of the $R$--symmetry 
SU$(\cN)$, 
satisfies the following infinitesimal 
transformation law 
under the superconformal group
\bea
\d\, \cO &=& - \x \, \cO 
+ \hf \,\o^{ab} M_{ab}  \, \cO  
+ {\rm i} \, \L_j{}^i \,
R_i{}^j \, \cO \non \\
&&- 2\left( q\, \s + \bar{q}\, \bar{\s} \right) 
\, \cO~.
\eea
Here $M_{a b}$ 
are the Lorentz generators,  
and $R_i{}^j$ are the generators of SU$(\cN)$.
The constant parameters $q$ and $\bar q$ 
determine the dimension  $(q+\bar q)$
and U(1) $R$--symmetry charge $(q-\bar q)$ 
of the superfield, respectively.

In $\cN=1$ superconformal theory, 
the supercurrent $J_{\a \ad}$ and 
the flavor current $L$ are primary superfields
with the superconformal transformations
\bea
\d J_{\a \ad}  &=& - \x\,J_{\a \ad} 
- 3\left(  
\s +  \bar{\s}  \right) J_{\a \ad} \non \\
&&+ ( \o_\a{}^\b  \d_\ad{}^\bd 
+ \bar{\o}_\ad{}^\bd  \d_\a{}^\b) J_{\b \bd}~,  \\
\d L  &=& -\x\,L    
 - 2\left(  \s  +  \bar{\s}  \right) L ~.
\eea
In $\cN=2$ superconformal theory, 
the supercurrent $\cJ$ and 
the flavor current $\cL_{ij}$ are primary superfields
with the superconformal transformations
\bea
\d \cJ  &=& -\x\,\cJ    
 - 2\left(  \s +  \bar{\s}  \right) \cJ ~,
\label{n=2sctrl}\\
\d \cL_{ij} &=& -\x \, \cL_{ij} 
- 2\left(  \s +  \bar{\s}  \right)\cL_{ij} \non \\
&&+\, 2 {\rm i}\, \L_{(i}{}^k \, \cL_{j)k}~. 
\label{n=2fcstl}
\eea 

Correlation functions of primary superfields,
$\langle \, \cO_1 (z_1)\,\cO_2 (z_2) \dots
\cO_n (z_n) \, \rangle$, involve some universal 
building blocks which we are going to describe briefly.
Associated with any two points $z_1$ and $z_2$ in superspace
are (anti-)chiral combinations
$x_{{\bar 1}2}$, $\q_{12}$ and ${\bar \q}_{12}\,$:
\bea
x^a_{{\bar 1}2} &=& - x^a_{2 {\bar 1}} =  x^a_{1-} - x^a_{2+} 
+2{\rm i}\, \q_{2 i}\, \s^a \,{\bar \q}_1^i~, \non \\
x^a_\pm & \equiv & x^a \pm {\rm i}\, \q_i \s^a {\bar \q}^i~;
 \non \\
\q_{12}&=& \q_1 - \q_2~, \qquad \quad {\bar \q}_{12} 
= {\bar \q}_1 - {\bar \q}_2~,
\eea  
which are invariant under Poincar\'e supersymmetry 
transformations (the notation `$x_{{\bar 1}2}$' indicates
that $x_{{\bar 1}2}$ is antichiral with respect to $z_1$
and chiral with respect to $z_2$)
but transforms semi-covariantly with respect 
to the superconformal group (see, e.g. \cite{KT}). 
In extended supersymmetry,
there exist primary superfields with isoindices,
and their correlation functions generically involve
a conformally covariant $\cN \times \cN$ unimodular 
matrix\footnote{We use the notation adopted  in \cite{wb,bk}.
When the spinor indices are not indicated explicitly,
the following matrix-like conventions are used \cite{osborn}:
$\j = (\j^\a)$, $\tilde{\j} = (\j_\a)$,
$\bar{\j} = ( \bar{\j}^\ad)$, 
$\tilde{\bar{\j}} = (\bar{\j}_\ad)$, 
$x = (x_{\a \ad})$, $\tilde{x} =(x^{\ad \a})$;
but $x^2 \equiv x^a x_a = - \hf \,{\rm tr}\, (\tilde{x} x)$,
and hence $\tilde{x}^{-1} = - x / x^2$.
} \cite{park2}
\bea
 \hat{u}_i{}^j (z_{12}) &=& \left( 
\frac{ x_{ {\bar 2} 1}{}^2 }{x_{ {\bar 1} 2}{}^2} 
\right)^{1/{\cN} } 
u_i{}^j (z_{12})~,  \non \\
 u_i{}^j (z_{12}) &=& \d_i{}^j - 4{\rm i}\;
\frac{ \q_{12\,i}x_{ {\bar 1} 2} {\bar \q}^j_{12} }
{x_{ {\bar 1} 2}{}^2} ~,
\label{hat-u}
\eea
with the basic properties
\bea
& \hat{u}^\dag (z_{12})~ \hat{u} (z_{12}) =  {\bf 1}~,& \non \\
& \hat{u}^{-1} (z_{12}) 
= \hat{u} (z_{21})~,& \non \\
& \det \; \hat{u} (z_{12}) = 1~, &
\label{unitary}
\eea
and the transformation rule
\bea
\d \hat{u}_i{}^j (z_{12}) &=& {\rm i}\,
\L_i{}^k (z_1) \hat{u}_k{}^j (z_{12}) \non \\
&-& 
{\rm i}\,\hat{u}_i{}^k (z_{12})\L_k{}^j (z_2) ~.
\eea

Given three superspace points $z_1,~z_2$ and $z_3$,
one can define superconformally covariant bosonic 
and fermionic variables $\bZ_1,~ \bZ_2$ and $\bZ_3$,
where $\bZ_1 = (\bX_1, ~ \Q^i_{1},~ 
{\bar \Q}_{1\,i})$ are \cite{osborn,park2}
\bea
\bX_1 & = &\tilde{x}_{1 \bar{2}}{}^{-1} 
\tilde{x}_{\bar{2} 3} \tilde{x}_{3 \bar{1}}{}^{-1} ~, \non \\
\tilde{\Q}_1^i & = & {\rm i}\, 
\left( \tilde{x}_{ \bar{2} 1}{}^{-1} {\bar \q}^i_{12} 
- \tilde{x}_{ \bar{3} 1}{}^{-1} {\bar \q}^i_{13} \right) ~,
\non \\
\bar{\bX}_1 &=& \bX^\dag_1 = \bX_1 -
4{\rm i}\, \tilde{\Q}^i_{1} \tilde{{\bar \Q}}_{1\, i}~, \non \\
\qquad
\tilde{ \bar{\Q}}_{1\,i}  &=& (\tilde{\Q}_1^i)^\dag  
\label{capZ}
\eea
and $\bZ_2,~\bZ_3$ are obtained from here by 
cyclically permuting indices.
These structures possess remarkably simple 
superconformal transformation rules:
\bea
\d \bX_{1\, \a \ad} & = & 
\left( \o_\a{}^\b (z_1) - \d_\a{}^\b \,\s (z_1) \right)
\bX_{1\, \b \ad} \non \\
&+& \bX_{1\, \a \bd}
\left( {\bar \o}^\bd{}_\ad (z_1)   
- \d^\bd{}_\ad  \,{\bar \s}(z_1)\right)~, \non \\
\d \Q^i_{1\, \a} & = & 
\o_\a{}^\b (z_1) \Q^i_{1\, \b} -
{\rm i} \Q^j_{1\, \a} \L_j{}^i (z_1) \non \\
&- &\frac{1}{\cN} \Big( (\cN - 2) \s (z_1) + 
2{\bar \s} (z_1) \Big)\Q^i_{1\, \a} 
\non
\eea
and turn out to be essential building blocks
for correlations functions of primary superfields.
The variables $\bZ$ with different labels are related to each
other, in particular:
\bea
\tilde{x}_{\bar{1} 3}\, \bX_3 \, \tilde{x}_{\bar{3} 1} &=&
- {\bar \bX}_1{}^{-1} ~, \non \\ 
\tilde{x}_{\bar{1} 3} \, \tilde{\Q}^i_3 \, u_i{}^j(z_{31})
&=& - \bX_1{}^{-1} \, \tilde{\Q}^j_1~. 
\label{difz}
\eea
With the aid of the matrices $u (z_{rs})$, $r,s = 1,2,3$, 
defined in (\ref{hat-u}),
one can construct unitary matrices $\hat{{\bf u}} (\bZ_s)$
\cite{park2},
in particular
\bea
\hat{{\bf u}} (\bZ_3) &=& \hat{u} (z_{31}) \hat{u} (z_{12})
\hat{u} (z_{23}) \label{greatu} \\
&=&
\left(
 \frac{ {\bar \bX}_3{}^2 }{ \bX_3{}^2 }\right)^{1 / \cN} 
\left( 
\d_i{}^j 
- 4{\rm i} \tilde{ \bar{\Q} }_{3\,i} \bX_3{}^{-1} 
\tilde{\Q}^j_3 \right) \non
\eea
transforming at $z_3$ only. Their properties are
\be
\hat{{\bf u}}^\dag (\bZ_3)= \hat{{\bf u}}^{-1} (\bZ_3)~,\qquad
\det  \hat{{\bf u}} (\bZ_3) = 1~.
\ee

The above general formalism has 
specific features in the case $\cN=2$ that is
of primary interest for us. Here
we have at our disposal 
the SU(2)--invariant tensors $\ve_{ij} = -\ve_{ji} $
and $\ve^{ij} = -\ve^{ji} $, normalized to
$\ve^{\underline{12}}= \ve_{\underline{21}} =1$.
They can be used
to raise and lower isoindices:
$C^i =  \ve^{ij} C_j \,$, 
$C_i =  \ve_{ij} C^j$.
{}For $\cN=2$, the condition of unimodularity of the matrix 
defined in (\ref{hat-u}) can be written as
\be 
 \hat{u}_{ji} (z_{21}) ~=~ -\hat{u}_{ij} (z_{12})~.
\label{unimod4}
\ee
The importance of this relation is that it implies that
the two-point function
\bea
A_{i_1 i_2} (z_1, z_2) & \equiv & 
\frac{ \hat{u}_{i_1 i_2} (z_{12})}
{ \left( x_{\bar{1} 2}{}^2   x_{\bar{2} 1}{}^2 \right)^\hf } \non \\
&=& - \frac{ \hat{u}_{i_2 i_1} (z_{21})}
{\left( x_{\bar{1} 2}{}^2   x_{\bar{2} 1}{}^2 \right)^\hf }
\eea
is analytic \cite{gikos} in $z_1$ and $z_2$ for $z_1 \neq z_2$,
\bea
D_{1\,\a (j_1} A_{i_1) i_2} (z_1, z_2) &=&0 ~, \non \\
{\bar D}_{1\,\ad (j_1} A_{i_1) i_2} (z_1, z_2) &=& 
0~.
\label{import111}
\eea
As we will see later, $A_{i_1 i_2} (z_1, z_2)$ is a building
block of correlation functions of analytic primary superfields
like the $\cN =2$ flavor currents.
{}For $\cN=2$, the fact that $\hat{{\bf u}} (\bZ_3)$ 
is unimodular and unitary, implies 
\bea
{\rm tr}\; \hat{{\bf u}}^\dag (\bZ_3) &=&
{\rm tr}\; \hat{{\bf u}} (\bZ_3)~, \non \\
 \hat{{\bf u}}^\dag_{ji} (\bZ_3) 
&=& - \hat{{\bf u}}_{ij} (\bZ_3)~.
\eea

\section{Correlation functions of \mbox{$\cN$} = 2 currents}

According to the general prescription of \cite{osborn,park2},
the two-point function of a primary superfield
$\cO_\cI$, which is a Lorentz scalar and transforms
in a representation $T$ of the $R$--symmetry group SU$(\cN)$,
with its conjugate $\bar{\cO}^\cJ$ reads
\bea
 \langle \cO_\cI (z_1)\;\bar{\cO}^\cJ (z_2)\rangle
&=& C_{\cO}\;\frac{ 
T_\cI{}^\cJ \Big( \hat{u}(z_{12}) \Big)}
{ (x_{\bar{1}2}{}^2)^{\bar q} (x_{\bar{2}1}{}^2)^q }~,
\non
\eea
where $ C_{\cO}$ is a normalization constant. 

{}For the $\cN=2$ supercurrent $\cJ$
and the flavor current $\cL^{\bar{a}}_{ij}$,
the above prescription gives
\bea
&& \langle \cJ (z_1)\;\cJ (z_2)\rangle 
 =  c_\cJ \; \frac{1}{ x_{ {\bar 1} 2}{}^2  
x_{{\bar 2} 1}{}^2}~,   \label{n=2sc,t-pf} \\
&&  \langle \cL^{\bar{a}_1}_{i_1 j_1} (z_1) 
\cL^{\bar{a}_2 \,i_2 j_2} (z_2)\rangle 
 =  2 c_\cL \,\d^{ \bar{a}_1 \bar{a}_2}\,  
\non \\ 
&& \qquad   \times 
\frac{
\hat{u}_{i_1}{}^{( i_2}(z_{12}) \;  \hat{u}_{j_1}{}^{ j_2)}(z_{12}) 
}
{  x_{ {\bar 1} 2}{}^2  x_{{\bar 2} 1}{}^2} ~. 
\label{n=2fcst-pf}
\eea
The relevant conservation equations prove
to be satisfied at $z_1 \neq z_2$,
\bea
D_1{}^{ij} \langle \cJ (z_1)\;\cJ (z_2) \rangle & =& 0~, \non \\
D_{1\, \a (k_1} \langle \cL_{i_1 j_1)} (z_1) 
\cL^{i_2 j_2} (z_2)\rangle &=&0~.
\eea 

According to the general prescription of 
\cite{osborn,park2}, the three-point function 
of primary superfields $\cO^{(1)}_{\cI_1}$,
$\cO^{(2)}_{\cI_2}$ and $\cO^{(3)}_{\cI_3}$
reads
\bea
&& \langle
\cO^{(1)}_{\cI_1} (z_1)\, \cO^{(2)}_{\cI_2} (z_2)\,
\cO^{(3)}_{\cI_3} (z_3)
\rangle \non \\
&& \qquad ~=~ 
\frac{ 
T^{(1)}{}_{\cI_1}{}^{\cJ_1} \left( \hat{u}(z_{13}) \right)
T^{(2)}{}_{\cI_2}{}^{\cJ_2} \left( \hat{u}(z_{23}) \right)
}
{ 
(x_{\bar{1}3}{}^2)^{\bar{q}_1} (x_{\bar{3}1}{}^2)^{q_1} 
(x_{\bar{2}3}{}^2)^{\bar{q}_2} (x_{\bar{3}2}{}^2)^{q_2}
}\; \non \\
&& \qquad \qquad  ~ \times ~ H_{\cJ_1 \cJ_2 \cI_3} (\bZ_3)~. \non
\eea 
Here $H_{\cJ_1 \cJ_2 \cI_3} (\bZ_3)$ transforms
as an isotensor at $z_3$ in the representations
$T^{(1)},~T^{(2)}$ and $T^{(3)}$ with respect 
to the indices $\cJ_1,~ \cJ_2$ and $\cI_3$, respectively,
and possesses the homogeneity property
\bea
&& H_{\cJ_1 \cJ_2 \cI_3} ( \D \bar{\D}\,\bX,
\D\, \Q, \bar{\D} {\bar \Q}) \non \\
&& \qquad =
\D^{2p} \bar{\D}^{2\bar{p}}
H_{\cJ_1 \cJ_2 \cI_3} ( \bX, \Q, {\bar \Q})~, \non \\
&& \qquad \qquad p- 2\bar{p} = \bar{q}_1 + \bar{q}_2 -q_3~,
\non \\
&& \qquad \qquad \bar{p} - 2p = q_1 + q_2 - \bar{q}_3~. \non
\eea
In general, the latter equation admits
a finite number of linearly independent solutions,
and this can be considerably reduced by taking into
account the symmetry properties, superfield conservation
equations and, of course, the superfield constraints
(such as chirality or analyticity \cite{gikos}). 
Below we shall present the most general 
expressions for three-point functions
of the $\cN=2$ supercurrent $\cJ$
and the flavor current $\cL^{\bar{a}}_{ij}$,
which are compatible with all physical
requirements. Details can be found in \cite{KT}.

The three-point function 
of the $\cN=2$ supercurrent is
\bea
&& \langle
\cJ (z_1)\, \cJ (z_2)\, \cJ (z_3)
\rangle   \label{sc3-pf}  \\
&& \qquad ~=~  \frac{1}{ 
x_{\bar{1}3}{}^2 x_{\bar{3}1}{}^2 
x_{\bar{2}3}{}^2 x_{\bar{3}2}{}^2} 
\non \\
&& \qquad ~ \times ~  \Bigg\{
A \, \Big(\frac{1}{ \bX_3{}^2} + \frac{1}{{\bar \bX}_3{}^2 }\Big) 
\non \\
&& \qquad ~+~ B\, \frac{ 
\Q_3^{\a \b} \bX_{3 \a \ad} \bX_{3 \b \bd} 
{\bar \Q}_3^{\ad \bd} }
{(\bX_3{}^2)^2}  \Bigg\}~,
\non
\eea
where
\bea
\Q_3^{\a \b} &=& \Q_3^{(\a \b)}~=~
\Q_3^{\a i} \, \Q_{3\,i}^\b~, \non \\ 
\bar{\Q}_3^{\ad \bd} &=& \bar{\Q}_3^{(\ad \bd)}~=~
\bar{\Q}_{3\,i}^\ad \, \bar{\Q}_3^{\ad i}~,
\eea
and  $A,~B $ are real parameters. The second 
structure is nilpotent and real.

The three-point function of the $\cN=2$ flavor current 
reads
\bea
&& \langle 
\cL^{\abar}_{i_1 j_1} (z_1)\,  \cL^{\bbar}_{i_2 j_2} (z_2)\,
\cL^{\cbar}_{i_3 j_3} (z_3)
\rangle \label{abc}  \\
&=& \frac{ 
\hat{u}_{i_1}{}^{k_1} (z_{13})  \hat{u}_{j_1}{}^{l_1} (z_{13}) 
\hat{u}_{i_2}{}^{k_2} (z_{23})  \hat{u}_{j_2}{}^{l_2} (z_{23})
}{
x_{ {\bar 3} 1}{}^2  x_{{\bar 1} 3}{}^2
x_{ {\bar 3} 2}{}^2  x_{{\bar 2} 3}{}^2 } 
\non \\
& \times &
f^{{\abar} {\bbar} {\cbar} } 
\left\{
\frac{ \ve_{ i_3(k_1 } \hat{{\bf u}}{}_{\,l_1)(l_2} (\bZ_3 )
\ve_{ k_2)j_3 } } { (\bX_3{}^2 \bar{\bX}_3{}^2 )^\hf} ~ 
+ ~ (i_3 \leftrightarrow j_3) 
\right\} \non
\eea
with $f^{\abar \bbar \cbar}=
f^{[\abar \bbar \cbar]}$ a completely antisymmetric real tensor 
being proportional to the structure constants of the flavor group.

{}For mixed correlation functions of the $\cN=2$
supercurrent and the flavor current, we get
\bea
&& \langle
\cJ (z_1) \; \cJ (z_2) \; \cL^{\abar}_{ij} (z_3)
\rangle = 0~, \label{zero} \\
&& \langle
\cL^{\abar}_{i_1 j_1} (z_1)\,  \cL^{\bbar}_{i_2 j_2} (z_2)\,
\cJ (z_3)
\rangle = d\;\d^{{\abar} {\bbar} } \; 
\label{l-l-j}  \\
& \times &  
\frac{ 
\hat{u}_{i_1}{}^{k_1} (z_{13})  \hat{u}_{j_1}{}^{l_1} (z_{13}) 
\hat{u}_{i_2}{}^{k_2} (z_{23})  \hat{u}_{j_2}{}^{l_2} (z_{23})
}{
x_{ {\bar 3} 1}{}^2  x_{{\bar 1} 3}{}^2
x_{ {\bar 3} 2}{}^2  x_{{\bar 2} 3}{}^2 } \non \\
& \times &
\frac{ \ve_{ k_2(k_1 } \hat{{\bf u}}{}_{\,l_1)l_2} (\bZ_3 )
+ \ve_{ l_2(k_1 } \hat{{\bf u}}{}_{\,l_1)k_2} (\bZ_3 )
} { (\bX_3{}^2 \bar{\bX}_3{}^2 )^\hf} ~, 
\non
\eea 
with $d$  a real parameter which can be 
related, via supersymmetric Ward identities, 
to the parameter $c_\cL$ in the two-point
function (\ref{n=2fcst-pf}),
\be
d ~=~ \frac{1}{4\p^2}\; c_\cL~.
\ee  
It is worth pointing out that eq. (\ref{zero})
is one of the important consequences of $\cN=2$
superconformal symmetry and has no direct analog 
in the $\cN=1$ case. In a generic $\cN=1$
superconformal theory with a flavor current
$L$, the correlation 
function $\langle J_{\a \ad}\, J_{\b \bd}\, L \rangle$
is not restricted by $\cN=1$
superconformal symmetry to vanish \cite{osborn}.

\section{Reduction to \mbox{$\cN$} = 1 superfields}
{}From the point of view of 
$\cN=1$ superconformal symmetry,
any $\cN=2$ primary superfield
consists of several $\cN=1$ primary superfields. 
Having computed the correlation functions
of $\cN=2$ primary superfields,
one can read off all correlators
of their $\cN=1$ superconformal components.
Since any $\cN=2$ superconformal theory 
is a particular $\cN=1$ superconformal theory,
one can then simply make use of $\cN=1$ 
superconformal Ward identities \cite{osborn}
to relate the coefficients of various correlators.

Using the explicit form (\ref{n=2sc,t-pf}) 
of the $\cN=2$ supercurrent two-point function,
one can read off the two-point functions of the 
$\cN=1$ primary superfields contained in 
$\cJ$, in particular\footnote{Here and below, 
all building blocks are 
expressed in $\cN=1$ superspace.}
\bea
\langle
J (z_1)\;J (z_2)
\rangle 
&=&  c_\cJ \; \frac{1}{ x_{ {\bar 1} 2}{}^2  
x_{{\bar 2} 1}{}^2}\,,  \label{n=1sct-pf} \\
\langle
J_{\a \ad}(z_1)\;J_{\b \bd} (z_2)
\rangle
&=&  \frac{64}{3}\, c_\cJ \;
\frac{ (x_{1\bar{2}})_{\a \bd}
(x_{2\bar{1}})_{\b \ad}}{( x_{ {\bar 1} 2}{}^2  
x_{{\bar 2} 1}{}^2)^2}\,. 
\non
\eea 
Similarly, the two-point function of the $\cN=1$
flavor current  
follows from (\ref{n=2fcst-pf})
\be
\langle
L^{\bar{a}_1} (z_1) \, L^{\bar{a}_2} (z_2)
\rangle  \;=\;
c_\cL 
\; \frac{\d^{\bar{a}_1 \,\bar{a}_2} }
{  x_{ {\bar 1} 2}{}^2  x_{{\bar 2} 1}{}^2} ~.
\ee

We now present several $\cN=1$ three-point functions
which are encoded in that of the $\cN=2$ supercurrent,
given by eq. (\ref{sc3-pf}). 
\bea
&& \langle
 J (z_1)\, J (z_2)\, J (z_3) \rangle  
= \frac{A}{ 
x_{\bar{1}3}{}^2 x_{\bar{3}1}{}^2 
x_{\bar{2}3}{}^2 x_{\bar{3}2}{}^2} \non \\
&& \qquad  \times 
\left(\frac{1}{ \bX_3{}^2} + \frac{1}{{\bar \bX}_3{}^2 }\right)~,
\eea
\bea
&& \langle
J (z_1)\, J (z_2) \,J_{\a \ad} (z_3) 
\rangle  = 
- \frac{1 }{12} (8 A - 3B)\; \non \\
&& \qquad  \times \frac{1}{ 
x_{\bar{1}3}{}^2 x_{\bar{3}1}{}^2 
x_{\bar{2}3}{}^2 x_{\bar{3}2}{}^2 } 
\non  \\
& & \qquad \times
\Bigg\{ 
\frac{ 2 (\bP_3 \cdot \bX_3) \bX_{3\, \a \ad} 
+ \bX_3{}^2 \bP_{3\, \a \ad}  }
{(\bX_3{}^2)^2 } \non \\ 
&& \qquad \qquad ~+~(\bX_3  \leftrightarrow  - \bar{\bX}_3)
\Bigg\}~, 
\label{JJ-sc}
\eea
\bea
&& \langle
J_{\a \ad} (z_1)\, J_{\b \bd} (z_2)\, J (z_3)
\rangle  = 
-\frac{4}{9}(8 A +3 B)\; \non \\
&& \qquad  \times  \frac{
(x_{1 \bar{3}})_{\a \dot{\g}}
(x_{3 \bar{1}})_{\g \dot{\a}}
(x_{2 \bar{3}})_{\b \dot{\d}}
(x_{3 \bar{2}})_{\d \dot{\b}}
}{ 
(x_{\bar{1}3}{}^2 x_{\bar{3}1}{}^2 
x_{\bar{2}3}{}^2 x_{\bar{3}2}{}^2)^2 } 
\non  \\
& & \qquad \times  
\Bigg\{
\frac{ 
\bX_3{}^{ \g \dot{\g} }  \bX_3{}^{ \d \dot{\d} }   }
{ (\bX_3{}^2)^3 } +\hf \,
\frac{ 
{\ve}^{\g \d}\; {\ve}^{ \dot{\g} \dot{\d} } }
{(\bX_3{}^2)^2 } \non \\
&& \qquad \qquad 
~+~(\bX_3  \leftrightarrow  - \bar{\bX}_3) 
\Bigg\}~,
\label{sc-sc-fcs}
\eea
with $\bP_a$ defined by  \cite{osborn}
\be
{\bar \bX}_a - \bX_a = {\rm i}\, \bP_a ~, \qquad 
\bP_a = 2\, \Q \, \s_a \, \bar \Q ~.
\ee
The most interesting correlator 
and by far the most laborious to compute is 
\bea
&& \langle
J_{\a \ad} (z_1)\, J_{\b \bd} (z_2)\, J_{\g \dot{\g}} (z_3)
\rangle  \label{final1} \\
&& \qquad = 
\frac{
(x_{1 \bar{3}})_{\a \dot{\s}}
(x_{3 \bar{1}})_{\s \dot{\a}}
(x_{2 \bar{3}})_{\b \dot{\d}}
(x_{3 \bar{2}})_{\d \dot{\b}}
}{ 
(x_{\bar{1}3}{}^2 x_{\bar{3}1}{}^2 
x_{\bar{2}3}{}^2 x_{\bar{3}2}{}^2)^2 }\; \non \\
&& \qquad  \times ~
H^{\dot{\s} \s, \dot{\d} \d}{}_{\g \dot{\g}} 
(\bX_3, \bar \bX_3)~, 
\non \\
&& H^{\dot{\s} \s, \dot{\d} \d }{}_{\g \dot{\g}} 
(\bX_3, \bar \bX_3)  =  
h^{\dot{\s} \s, \dot{\d} \d}{}_{\g \dot{\g}} 
(\bX_3, \bar \bX_3) \non \\
&& \qquad \qquad + ~
h^{ \dot{\d} \d, \dot{\s} \s }{}_{\g \dot{\g}} 
(- \bar \bX_3, - \bX_3) ~,
\non
\eea
where 
\bea
&& h^{abc}(\bX, \bar \bX)  \equiv  
-\frac{1}{8}\,(\s^a)_{\a \ad}\,(\s^a)_{\b \bd}\,
(\tilde{\s}^c)^{\dot{\g} \g}\; \non \\
&& \times h^{\ad \a, \bd \b}{}_{\g \dot{\g}} 
(\bX, \bar \bX) 
\label{final2}\\
&& = - \frac{16}{27} \,(26A-\frac{9}{4}B) 
\; \frac{ {\rm i}  } {(\bX^2)^2}\; \non \\
&& \times
\Big(\bX^a \eta^{bc} + \bX^b \eta^{ac} - \bX^c \eta^{ab} 
+{\rm i}\, \ve^{abcd} \bX_d \Big) \non \\
&&- \frac{8}{27}\, (8A-9B)\;
\frac{ {\rm 1}  }  
{(\bX^2)^3}\; \non \\
&& \times \Bigg\{
2\Big(\bX^{a} \bP^{b} 
+ \bX^{b} \bP^{a} \Big) \bX^c \non \\
&&-3\bX^a \bX^b \Big( \bP^c
+2\frac{(\bP \cdot \bX)}{\bX^2} \bX^c \Big) \non  \\
&& - (\bP \cdot \bX)\Big( 
3( \bX^a \eta^{bc} + \bX^b \eta^{ac} ) -2 \bX^c \eta^{ab} \Big)
\non \\
&&+\hf \bX^2 \Big( \bP^a \eta^{bc} + \bP^b \eta^{ac} 
 + \bP^c \eta^{ab} \Big) \Bigg\}~. \non
\eea 
Our final relations (\ref{final1}) and (\ref{final2})
perfectly agree with the general structure of the three-point
function of the supercurrent in $\cN=1$ superconformal 
field theory \cite{osborn}.

Using the results of \cite{osborn}, 
one may express
$A$ and $B$ in terms of the anomaly coefficients \cite{Anselmietal} 
\bea
a &=& \frac{1}{24}(5n_V +n_H)~, \non \\
c &=& \frac{1}{12}(2n_V +n_H)~,
\eea 
where $n_V$ and $n_H$ denote the number of free $\cN = 2$
vector multiplets and hypermultiplets, respectivley.
We get\footnote{Our definition of the $\cN=1$ supercurrent
corresponds to that adopted in \cite{bk} and differs in sign
from Osborn's convention \cite{osborn}.}  
\bea
A &=& \frac{3}{64 \p^6}(4a - 3c)~, \non \\ 
B &=& \frac{1}{8 \p^6}(4a - 5c)~.
\eea

In $\cN=1$ supersymmetry, a superconformal Ward identity
relates the coefficient in the two-point function of the 
supercurrent (\ref{n=1sct-pf}) to the anomaly coefficient
$c$ as follows \cite{osborn}
\be
c_\cJ = \frac{3}{8\p^4} \,c~.
\ee
In terms of the coefficients $A$ and $B$ this relation reads
\be
\frac{2}{\p^2}\,c_\cJ =  8A-3B ~.
\ee

${}$Let us turn to the three-point function 
of the $\cN=2$ flavor current given by eq.
(\ref{abc}). 
${}$From it one reads off
the three-point function of the $\cN=1$ flavor current
\bea
&& \langle
L^{\bar{a}} (z_1)\, L^{\bar{b}} (z_2)\, 
L^{\bar{c}} (z_3)
\rangle  \label{n=2t-pffcs2} \\
&& \qquad = 
\frac{1}{4}\,f^{\bar{a} \bar{b} \bar{c}} 
\frac{{\rm i}}{ 
x_{\bar{1}3}{}^2 x_{\bar{3}1}{}^2 
x_{\bar{2}3}{}^2 x_{\bar{3}2}{}^2} \non \\
&& \qquad \times ~ \left(  \frac{1}{{\bar \bX}_3{}^2 }
- \frac{1}{ \bX_3{}^2}
\right) ~.
\non
\eea
It is worth noting that the Ward identities allow one to represent
$f^{\bar{a} \bar{b} \bar{c}}$
as a product of $c_\cL$ and the structure constants of
the flavor symmetry group, see \cite{osborn} for more details.

In $\cN=1$ superconformal 
field theory, the three-point 
function of the flavor current superfield $L$
contains, in general, two linearly independent forms \cite{osborn}:
\bea
&& \langle
 L^{\bar{a}} (z_1)\, L^{\bar{b}} (z_2)\, 
L^{\bar{c}} (z_3)
\rangle \non \\
& & \qquad = ~
\frac{1}{ 
x_{\bar{1}3}{}^2 x_{\bar{3}1}{}^2 
x_{\bar{2}3}{}^2 x_{\bar{3}2}{}^2} \non \\ 
&& \qquad  \times ~ \Bigg\{
{\rm i}\, f^{[\bar{a} \bar{b} \bar{c}]} 
\left(\frac{1}{ \bX_3{}^2}-  
\frac{1}{{\bar \bX}_3{}^2 } \right) \non \\
&& \qquad \quad  ~+~  d^{(\bar{a} \bar{b} \bar{c})} 
\left(\frac{1}{ \bX_3{}^2}+  
\frac{1}{{\bar \bX}_3{}^2 } \right)\Bigg\}~.
\non 
\eea
The second term, involving a completely symmetric
group tensor $d^{\bar{a} \bar{b} \bar{c}}$,
reflects the presence of chiral anomalies in the theory. 
The field-theoretic origin of this term is due to the
fact that the $\cN=1$ conservation equation
$\bar{D}^2\, L = D^2\, L = 0$ admits a non-trivial deformation
\bea
\bar{D}^2\, \langle L^{ \bar{a} } \rangle ~\propto ~
d^{\bar{a} \bar{b} \bar{c}}\;
W^{\bar{b}\, \a} \, W^{\bar{c}}_\a \non
\eea
when the chiral flavor current is coupled 
to a background vector multiplet.  
Eq. (\ref{n=2t-pffcs2}) tells us that the
flavor currents are anomaly-free in $\cN=2$
superconformal theory. This agrees 
with the facts that (i) $\cN=2$ super Yang-Mills
models are non-chiral; (ii)
the $\cN=2$ conservation
equation (\ref{scc2f}) 
does not possess 
non-trivial deformations. 
\vspace{7mm}

\noindent
{\bf Acknowledgements} \hfill\break 
Support from DFG, the German National Science Foundation,
from GIF, the German-Israeli foundation 
for Scientific Research and from the EEC under TMR
contract ERBFMRX-CT96-0045 is gratefully acknowledged.
This work was also supported in part
by the NATO collaborative research grant PST.CLG 974965,
by the RFBR grant No. 99-02-16617, by the INTAS grant
No. 96-0308 and  by the DFG-RFBR grant No. 99-02-04022.


\begin{thebibliography}{99}


\bibitem{KT}
S.M.~Kuzenko and S.~Theisen,
Class.\ Quant.\ Grav.\  {\bf 17} (2000) 665,
hep-th/9907107.

\bibitem{osborn} H. Osborn, Ann. Phys. {\bf 272} (1999)
243, hep-th/9808041.

\bibitem{park2} J-H. Park, Nucl. Phys. {\bf B539}
(1999) 599, hep-th/9807186;
Nucl.\ Phys.\  {\bf B559} (1999) 455, hep-th/9903230. 

\bibitem{maldacena}
J.~Maldacena,
Adv. Theor. Math. Phys. {\bf 2} (1998) 231,
hep-th/9711200.

\bibitem{aharonyetal} O.~Aharony, S.S.~Gubser, J.~Maldacena, 
H.~Ooguri and Y.~Oz, Phys. Rep. {\bf 323} (2000) 183,
hep-th/9905111.

\bibitem{fz} S. Ferrara and B. Zumino,
Nucl. Phys. {\bf 87B} (1975) 207.

\bibitem{WZ} J. Wess and B. Zumino, 
Nucl. Phys. {\bf B78} (1974) 1;
S. Ferrara and B. Zumino, 
Nucl. Phys. {\bf B79} (1974) 413;
A. Salam and J. Strathdee, 
Phys. Rev. {\bf D11} (1975) 1521.

\bibitem{sohnius} M.F. Sohnius, Phys. Lett. {\bf 81B} (1979) 8.

\bibitem{Mez} L. Mezincescu, {\it On the superfield formulation 
of O(2) supersummetry}, JINR preprint P2-12572, 1979 (unpublished).

\bibitem{hst} 
P. Howe, K.S. Stelle and P.K. Townsend,
Nucl. Phys. {\bf B192} (1981) 332.

\bibitem{wb} J. Wess and J. Bagger, 
{\it Supersymmetry and Supergravity}, 2nd ed.
(Princeton University Press, 1992).

\bibitem{bk} I.L. Buchbinder and S.M. Kuzenko,
{\it Ideas and Methods of Supersymmetry and 
Supergravity}, 2nd ed. (IOP, 1998). 

\bibitem{gikos} A. Galperin, E. Ivanov, S. Kalitzin, V. Ogievetsky 
and E. Sokatchev, Class. Quantum Grav. {\bf 1} (1984) 469.

\bibitem{Anselmietal} 
D. Anselmi, D.Z. Freedman, M.T. Grisaru and A.A. Johansen,
Nucl. Phys. {\bf B526} (1998) 543, hep-th/9708042.
\end{thebibliography}
\end{document}